\documentclass[aps,pra,twocolumn,amssymb,amsfonts,floatfix,showpacs,superscriptaddress]{revtex4}

\usepackage{graphicx}% Include figure files
\usepackage{dcolumn}% Align table columns on decimal point
\usepackage{bm}% bold math
\usepackage{mathbbol}
\usepackage{amsmath,amssymb,amsthm}
\usepackage{times}
\usepackage{graphicx}
\usepackage{bm}
\usepackage[dvips]{color}

\begin{document}

\title{Quench Dynamics of Isolated Many-Body Quantum Systems}
\author{E. J. Torres-Herrera}
\affiliation{Department of Physics, Yeshiva University, New York, New York 10016, USA}
\author{Lea F. Santos}
\affiliation{Department of Physics, Yeshiva University, New York, New York 10016, USA}

\begin{abstract}
We study isolated quantum systems with two-body interactions after a quench. In these systems, the energy shell is a Gaussian of width $\sigma$, and it gives the maximum possible spreading of the energy distribution of the initial states. When the distribution achieves this shape, the fidelity decay can be Gaussian until saturation. This establishes a lower bound for the fidelity decay in realistic systems.  An ultimate bound for systems with many-body interactions is also derived based on the analysis of full random matrices. We find excellent agreement between numerical and analytical results.  We also provide the conditions under which the short-time dynamics of few-body observables is controlled by $\sigma$.  The analyses are developed for systems, initial states, and observables accessible to experiments.
\end{abstract}

\pacs{75.10.Jm, 05.45.Mt, 05.70.Ln, 72.25.Rb}

\maketitle

%%%%%%%%%%%%%%% INTRODUCTION %%%%%%%%%%%%%%%%%%
\section{INTRODUCTION}
Nonequilibrium quantum physics is much less understood than equilibrium quantum physics. Advances for the first will impact fields as diverse as condensed matter physics, molecular dynamics, and cosmology. In this paper, we focus on a crucial aspect of this vast subject, namely the unitary dynamics of isolated many-body quantum systems initially far from equilibrium. This topic has gained enormous attention due to recent experiments with cold atoms in optical lattices~\cite{experimentRef,Trotzky2008,Trotzky2012,FukuharaPhys,Fukuhara}, where coherent evolution can be studied for long times. Knowing the maximum speed at which many-body systems can evolve~\cite{Bhattacharyya1983,Pfeifer1993,GiovannettiALL} is also central for the development of algorithms for quantum optimal control~\cite{Caneva2009}.

In experiments with optical lattices, the evolution of the system initiates after changing instantaneously (quenching) a certain initial Hamiltonian $\hat{H}_I$ to a new final Hamiltonian $\hat{H}_F$. In this context there is evidence that the relaxation dynamics shows a power law behavior in some disordered systems~\cite{Khatami2012}, noninteracting integrable systems~\cite{Gramsch2012,HeSantos2013}, and close to critical points~\cite{Venuti2010,criticalRef}.
Here we study deterministic (clean) quantum systems with interaction. For different final Hamiltonians, we identify features that lead to similar relaxation processes and can therefore contribute to a long aspired universal picture.

In 1984 Peres proposed that the notion of irreversibility in the quantum domain could be explained in terms of sensitivity of the quantum states to perturbations added to the Hamiltonian~\cite{Peres1984}.  Such sensitivity is quantified via fidelity (Loschmidt echo) between states evolved under an unperturbed Hamiltonian and states evolved by a perturbed Hamiltonian~\cite{Gorin2006,LoschRef}. Several studies confirmed Peres's expectations and showed that the fidelity decayed exponentially when the perturbation induced chaos~\cite{expRef,Cerruti2002,Flambaum2001ab,Weinstein2003}. However,  an exponential behavior was observed also in integrable systems when the initial state was sufficiently delocalized in the energy representation~\cite{Emerson2002,Santos2012PRL,Santos2012PRE}. This suggests that, rather than a chaotic Hamiltonian, the main cause for a fast statistical relaxation must be the chaotic structure of the initial state with respect to the Hamiltonian that dictates its evolution.

To establish a relationship between the level of delocalization of the initial state and the characteristics of the relaxation dynamics after a quench, we employ concepts  from  many-body quantum chaos. We associate the unperturbed Hamiltonian with  $\hat{H}_I$ and the perturbed Hamiltonian with $\hat{H}_F$. Our initial state $|\text{ini} \rangle$ is one of the eigenstates $ |n \rangle$ of $\hat{H}_I$. Its projection on the eigenstates $|\psi_{\alpha} \rangle$ of $\hat{H}_F$ leads to $ | \text{ini}  \rangle= \sum_{\alpha} C_{\alpha}^{\text{ini}} |\psi_{\alpha}  \rangle$, and its energy is given by ${E_{\text{ini}} = \langle \text{ini} |\hat{H}_F | \text{ini} \rangle = \sum_{\alpha} |C_{\alpha}^{\text{ini}}|^2 E_{\alpha} }$. The distribution $P^{\text{ini}}_{\alpha}$ of the components $|C_{\alpha}^{\text{ini}}|^2$ in the eigenvalues $E_{\alpha}$ is known as local density of states (LDOS)~\cite{Flambaum2000}. For $E_{\text{ini}}$ close to the middle of the spectrum, as the perturbation increases from zero, this distribution broadens from a delta function to a Lorentzian form and eventually approaches a Gaussian of width $\sigma_{\text{ini}}$, identified as the energy shell~\cite{noteShell,ZelevinskyRep1996,Flambaum1997,Flambaum2000,Flambaum2001ab,Santos2012PRL,Santos2012PRE,Torres2013,noteGauss}. Contrary to full random matrices, where the states are fully delocalized, in realistic systems with few-body interactions, the energy shell gives the maximum possible spreading of the LDOS. 

LDOS is a key concept in nuclear physics~\cite{ZelevinskyRep1996}, where it is measured experimentally. It is likely to gain a especial role also in quantum thermodynamics, since it is related with the probability distribution of work~\cite{work}

In this paper, we investigate LDOS, the fidelity decay, and the short-time dynamics of few-body observables in the limit of strong perturbation. The fidelity corresponds to the Fourier transform of the LDOS. It therefore decays exponentially when $P^{\text{ini}}_{\alpha}$ is Lorentzian~\cite{expRef,Cerruti2002,Flambaum2001ab,Weinstein2003,Emerson2002}, although at very short-times it behaves as $1 - \sigma_{\text{ini}}^2 t^2$, as expected from perturbation theory~\cite{Flambaum2001ab,Cerruti2002}. Interestingly, consensus has been reached that even when $P^{\text{ini}}_{\alpha}$ was Gaussian, a Gaussian decay would occur for a certain time and then necessarily switch to exponential before saturation~\cite{Flambaum2001ab,Santos2012PRL,Santos2012PRE,Castaneda}. Here, we show that when the initial state fills the energy shell substantially, the Gaussian expression for the fidelity, $\exp (-\sigma_{\text{ini}}^2 t^2)$, {\em can persist until saturation}, independent of the regime (integrable or chaotic) of $\hat{H}_F$. This is illustrated numerically for initial states that can be prepared in experiments with cold atoms, but can hold for any $|\text{ini} \rangle$ where $P^{\text{ini}}_{\alpha}$ has a Gaussian shape~\cite{preparation,preparationTorres}. The absence of an exponential decay also has been noticed in recent works~\cite{Genway} for models and initial states different from the ones we consider.

The Gaussian expression sets the lower bound for the fastest fidelity decay in realistic systems with two-body interactions that are strongly perturbed.
We also discuss the scenario where $\hat{H}_F$ is a full random matrix~\cite{Brody1981}. Although less realistic, the latter sets the ultimate bound for the fidelity decay in many-body quantum systems after a quench.

The analysis of the dynamics of few-body observables is more challenging and at the same time indispensable to establishing a connection with current experiments. We provide examples of observables that although evolving under entirely different Hamiltonians show almost identical short-time dynamics. We identify the conditions for such {\em general behavior}.

%%%%%%%%%%%%%%%%%%% MODEL%%%%%%%%%%%%%%%%%%%%%%%%%
\section{MODEL}
We consider a one-dimensional lattice of interacting spins 1/2 with open boundaries and an even number $L$ of sites. The Hamiltonian has nearest-neighbor (NN) and possibly also next-nearest-neighbor (NNN) couplings:
\begin{align}
  \hat{H} = &\hat{H}_{NN} + \lambda \hat{H}_{NNN} ,
 \label{ham} \nonumber\\
 \hat{H}_{NN} = &\sum_{i=1}^{L-1} J \left(\hat{S}_i^x \hat{S}_{i+1}^x + \hat{S}_i^y \hat{S}_{i+1}^y +\Delta \hat{S}_i^z \hat{S}_{i+1}^z \right) ,
 \\
\hat{H}_{NNN} =& \sum_{i=1}^{L-2} J \left(\hat{S}_i^x \hat{S}_{i+2}^x + \hat{S}_i^y \hat{S}_{i+2}^y +\Delta \hat{S}_i^z \hat{S}_{i+2}^z \right) .
\nonumber 
\end{align}
This is a prototype many-body quantum model, simulated in optical lattices~\cite{Trotzky2008,Trotzky2012,FukuharaPhys,Fukuhara} and also mappable onto systems of spinless fermions or hardcore bosons~\cite{Jordan1928}. $\hat{S}^{x,y,z}_i$ are the spin operators on site $i$; $\hbar=1$. The coupling strength $J$, the anisotropy $\Delta$, and the ratio $\lambda$ between NNN and NN exchanges are positive; $\hat{S}_i^x \hat{S}_{i+1 (i+2)}^x + \hat{S}_i^y \hat{S}_{i+1 (i+2)}^y$ 
is the flip-flop term and $\hat{S}_i^z \hat{S}_{i+1 (i+2)}^z $
is the Ising interaction. The Hamiltonian conserves total spin in the $z$ direction, $[\hat{H},{\cal \hat{S}}^z]=0$, where ${\cal \hat{S}}^z = \sum_{i=1}^L \hat{S}_i^z$. 

The noninteracting XX model ($\Delta = \lambda = 0$) is trivially solvable. The interacting XXZ case ($\Delta \neq 0$, $\lambda =0$) 
is solved with the Bethe ansatz ~\cite{Bethe1931}. The system undergoes a crossover to the chaotic regime~\cite{Brody1981} as $\lambda $ increases~\cite{Gubin2012}. 

We investigate the dynamics of the system for the following choices of parameters for $\hat{H}_F$: 
\begin{enumerate}
\itemsep-0.4em
\item[(1)] Integrable isotropic Hamiltonian, $\hat{H}_{\Delta=1,\lambda=0}$.

\item[(2)]  Integrable anisotropic Hamiltonian, $\hat{H}_{\Delta=0.5,\lambda=0}$.

\item[(3)]  Weakly chaotic isotropic Hamiltonian, $\hat{H}_{\Delta=1,\lambda=0.4}$.

\item[(4)]  Strongly chaotic isotropic Hamiltonian, $\hat{H}_{\Delta=1,\lambda=1}$.

\item[(5)]  Strongly chaotic anisotropic Hamiltonian, $\hat{H}_{\Delta=0.5,\lambda=1}$.
\end{enumerate}

Independent of the regime, the density of states of the five Hamiltonians above has a Gaussian shape, as typical of systems with two-body interactions~\cite{ZelevinskyRep1996,Izrailev1990,Kota2001}. This implies that the majority of the states concentrate in the middle of the spectrum. This is the region where the eigenstates are expected to reach their highest level of delocalization.
\vskip -0.5 cm
\begin{table}[h]
\caption{Energy of $|\text{ini}\rangle$ and width of its energy distribution.}
\begin{center}
\begin{tabular}{ccc}
\hline 
\hline 
  &    $E_{\text{ini}}$  & $\sigma_{\text{ini}}$ \\ [0.1 cm]
\hline  
$ |\rm{NS}\rangle$  & \hspace{0.2 cm} $ 
 \frac{\displaystyle J\Delta}{\displaystyle 4}  [ -(L-1) + (L-2)\lambda ]$ &   $\frac{\displaystyle J}{\displaystyle 2} \sqrt{L-1}$  \\  [0.2 cm]
 
$ |\rm{PS}\rangle$   & $ - \frac{\displaystyle J\Delta}{\displaystyle 4}  [ 1 + (L-2)\lambda ]$ &  $\frac{\displaystyle J}{\displaystyle 2} \sqrt{ \frac{\displaystyle L}{\displaystyle 2} + (L-2)\lambda^2}$ \\  [0.2 cm]

$|\rm{DW}\rangle$   & \hspace{0.2 cm} $\frac{\displaystyle J\Delta}{\displaystyle 4} [(L-3) + (L-6)\lambda]$ & $ \frac{\displaystyle J}{\displaystyle 2} \sqrt{1+2\lambda^2}$    \\ [0.2 cm]
\hline
\hline 
\end{tabular}
\end{center}
\label{table:initial}
\end{table}
\vskip -0.5 cm
%

%%%%%%%%%%%%%%%% INITIAL STATE %%%%%%%%%%%%%%%%%%%%%%%%%

\section{INITIAL STATE}. 
Our analysis focuses on  $|\text{ini} \rangle$'s where each site has a spin either pointing up or down  in the $z$ direction~\cite{Zangara2013,Santos2011,Pozsgay}. The chosen states enhance the effects of $\Delta$ or $\lambda$:
\begin{enumerate}
\itemsep-0.1em
\item[(1)] N\'eel state, $ |\rm{NS}\rangle= |  \downarrow \uparrow \downarrow \uparrow \ldots  \downarrow \uparrow \downarrow \uparrow  \rangle$, 

\item[(2)] Pairs of parallel spins, $ |\rm{PS}\rangle=| \downarrow \uparrow  \uparrow  \downarrow \downarrow \uparrow  \uparrow \downarrow \downarrow \ldots   \rangle$. 

\item[(3)] Sharp domain wall, $|\rm{DW}\rangle = | \uparrow \uparrow \uparrow \ldots \downarrow \downarrow \downarrow \rangle$, 
\end{enumerate}

They belong to the same subspace with ${\cal S}^z =0$ and dimension ${\cal D} = L!/(L/2)!^2$.
These states can be prepared in optical lattices: $|\rm{DW}\rangle$ requires a magnetic field gradient~\cite{Weld2009} and $|\rm{NS}\rangle$ was used in~\cite{Trotzky2008,Koetsier2008,Trotzky2012,Mathy2012}. 
Our $\hat{H}_I$ therefore coincides with the Ising part of (\ref{ham}). Its eigenstates are referred to as site-basis vectors.
The final Hamiltonian is written in this basis. $\hat{H}_F$ is in the nonperturbative regime, since the off-diagonal elements are much larger than the average level spacing.

\section{ENERGY SHELL} 
The energy shell is a Gaussian centered at $E_{\text{ini}}$ of width 
\begin{eqnarray}
\sigma_{\text{ini}} &=& \sqrt{\sum_{\alpha} |C_{\alpha}^{\text{ini}} |^2 (E_{\alpha} - E_{\text{ini}})^2}
=\sqrt{\sum_{n \neq \text{ini}} |\langle n |\hat{H}_F | \text{ini}\rangle |^2 } \nonumber \\
&=& \frac{J}{2} \sqrt{M_1 + \lambda^2 M_2} .
\label{deltaE}
\end{eqnarray}

The second equality above shows that $\sigma_{\text{ini}}$ can be obtained from  $\hat{H}_F$ before diagonalization.
\begin{figure}[htb]
\centering
\includegraphics*[width=0.45\textwidth]{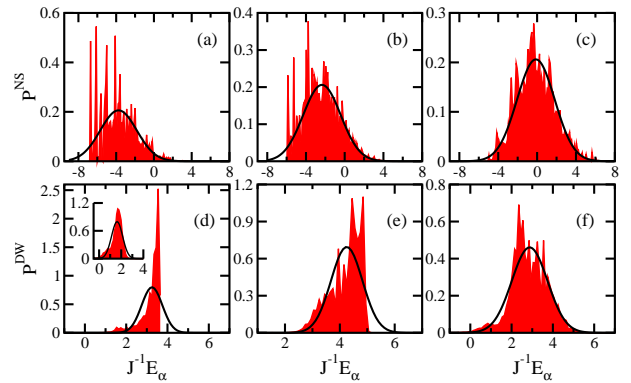}
\caption{(Color online) Distribution of  $|C_{\alpha}^{\text{ini}} |^2$ in $E_{\alpha}$ ({\em i.e.} LDOS) for 
$ |\rm{NS}\rangle$ (top) and $ |\rm{DW}\rangle$(bottom); 
$L=16$. The Hamiltonians are:  $\hat{H}_{\Delta=1,\lambda=0}$  (a, d); $\hat{H}_{\Delta=1,\lambda=0.4}$ (b, e);  $\hat{H}_{\Delta=0.5,\lambda=1}$ (c, f); and $\hat{H}_{\Delta=0.5,\lambda=0}$ [inset of (d)]. The solid line is the energy shell.}
\label{fig:shell}	
\end{figure}
The last equality holds for the models and initial states studied here. Our $\sigma_{\text{ini}}$ does not depend on the anisotropy parameter. The connectivity $M_1$ ($M_2$) corresponds to the number of states directly coupled to $|\text{ini}\rangle$ via the NN (NNN) flip-flop term. The total connectivity $M$ of any site-basis vector is low, $M\!=\!M_1 + M_2 \propto L \ll {\cal D}$. 

For $ |\rm{NS}\rangle$, the five Hamiltonians considered lead to the same $\sigma_{\text{ini}}$ (see Table~\ref{table:initial}), since $M_2=0$.
For $ |\rm{PS}\rangle$ and $|\rm{DW}\rangle$ only Hamiltonians with the same $\lambda$ give the same $\sigma_{\text{ini}}$. The domain wall is directly coupled to only one (three) state(s) when $\hat{H}_F$ is integrable (chaotic), independent of $L$. It has the smallest $\sigma_{\text{ini}}$ among the states investigated. 

Figure~\ref{fig:shell} displays the energy distribution of $ |\rm{NS}\rangle$ and $|\rm{DW}\rangle$ (see $ |\rm{PS}\rangle$ in Refs.~\cite{Zangara2013,preparation}).  As visible, $P^{\text{ini}}_{\alpha}$ depends on both $|\text{ini} \rangle $ and the $\hat{H}_F$ that evolves it. This dependence is reflected also in the relaxation dynamics of the system.

For the N\'eel state, $\sigma_{\text{ini}}$ is always the same, but the shell gets better filled as $\lambda$ increases from zero and $\Delta$ decreases [Fig.~\ref{fig:shell}  from (a) to (c)], since its energy is brought closer to the middle of the spectrum (cf. Table~\ref{table:initial}), where the density of states is larger. For $|\rm{DW}\rangle$, $\Delta$ plays a major role [cf. main panel and inset of Fig.~\ref{fig:shell} (d)]. At the critical point ($\Delta=1$) or above it, this state approaches the right edge of the spectrum. In this region, $|\rm{DW}\rangle$ and the few states directly coupled to it are more localized. As a result, in addition to the narrow energy shell, the latter is also poorly filled. $|\rm{DW}\rangle$ should therefore decay very slowly when $\Delta \geq 1$ \cite{Zangara2013,Santos2011,DWstudies}. For $|\rm{PS}\rangle$ the worst filling occurs for $\Delta=1$ and $\lambda=1$, since this combination pushes the distribution to the edge of the spectrum (cf. Table~\ref{table:initial}).

%%%%%%%%%%%%%%%% FIDELITY %%%%%%%%%%%%%%%%%%%%%%%%%
 
\section{FIDELITY} 
The probability of finding $|\text{ini} \rangle$ at time $t$ defines the fidelity. The latter corresponds to the Fourier transform of $|C_{\alpha}^{\text{ini}} |^2$.
When the energy distribution of $|\text{ini} \rangle$ fills the energy shell, $|C_{\alpha}^{\text{ini}} |^2$ is replaced by a Gaussian of $\sigma_{\text{ini}}$ (\ref{deltaE}) \cite{Flambaum2000,Flambaum2001ab}, so
\begin{eqnarray}
&&F(t) \equiv |\langle \text{ini} | e^{-i \hat{H}_F t} |\text{ini} \rangle |^2 \! = \! \left|\sum_{\alpha} |C_{\alpha}^{\text{ini}} |^2 e^{-i E_{\alpha} t}  \right|^2 
\nonumber \\
&& \approx \left| \frac{1}{ \sqrt{ 2 \pi \sigma^2_{\text{ini}} }  } \int_{-\infty}^{\infty} \! e^{-\frac{(E-E_{\text{ini}})^2}{ 2 \sigma^2_{\text{ini}} }  } \!e^{-i E t} dE
\right|^2 \!=\!  e^{-\sigma_{\text{ini}}^2 t^2} .
\label{eq:fidelity}
\end{eqnarray}
Above, the sum becomes an integral because ${\cal D}$ is large \cite{noteVenuti}. 
\begin{figure}[htb!]
\centering
\includegraphics[width=0.45\textwidth]{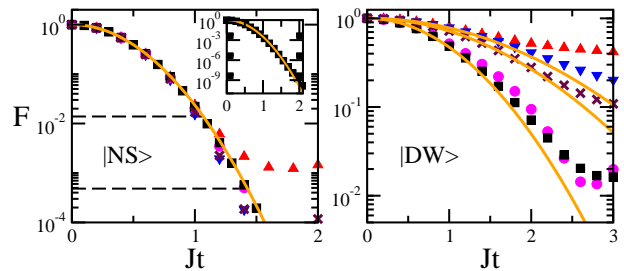}
\caption{(Color online) Fidelity decay for the final Hamiltonians:  $\hat{H}_{\Delta=1,\lambda=0}$ (up triangle),
$\hat{H}_{\Delta=0.5,\lambda=0}$ (down triangle), 
$\hat{H}_{\Delta=1,\lambda=0.4}$ (cross),
$\hat{H}_{\Delta=1,\lambda=1}$ (circle), and
$\hat{H}_{\Delta=0.5,\lambda=1}$ (square); $L=16$. [Inset: $\hat{H}_{\Delta=0.5,\lambda=1}$, $L\!=\!24$].  Solid lines are for Eq.~(\ref{eq:fidelity}). Dashed lines give $\overline{F}=\text{IPR}_{|\rm{NS}\rangle}^{-1}$: $\hat{H}_{\Delta=1,\lambda=0}$ (top) and $\hat{H}_{\Delta=0.5,\lambda=1}$ (bottom).}
\label{fig:fidelity}
\end{figure}
As seen in Fig.\ref{fig:fidelity}, $F(t)$ for $|\rm{NS}\rangle$ is similar for the five Hamiltonians, since $\sigma_{\text{ini}}$ is equal for all of them. The agreement between the Gaussian expression (\ref{eq:fidelity}) and the numerical results is excellent and holds until saturation (shown with dashed lines). This is not an artifact of small system sizes, as confirmed by the inset for larger $L$. This behavior contrasts  previous studies, where $F(t)$ was Gaussian for some time and then exponential before saturation~\cite{expRef,Cerruti2002,Flambaum2001ab,Weinstein2003,Emerson2002,Santos2012PRL,Santos2012PRE,Castaneda}. 

For $|\rm{DW}\rangle$ (Fig.~\ref{fig:fidelity}) and $|\rm{PS}\rangle$ (not shown), $F(t)$ coincides only for $\hat{H}_F$ with equal NNN flip-flop strength, since $\sigma_{\text{ini}}$ now depends on $\lambda$. The fidelity decay for $|\rm{DW}\rangle$ is slow due to its narrow energy distribution.  For both, $|\rm{DW}\rangle$ and $|\rm{PS}\rangle$, the filling of the energy shell is poorer when $\Delta=1$, so the Gaussian expression holds for longer in the anisotropic case. The filling is also overall worse than for $ |\rm{NS}\rangle$, which explains the transition of $F(t)$ to an exponential decay as $t$ increases. Identifying the critical time for this transition is far from trivial and strongly dependent on the initial state. It was discussed in~\cite{Flambaum2001ab} and estimates were made in Ref.~\cite{Castaneda}. 

The fidelity eventually saturates to its infinite time average $\overline{F}\!=\!\sum_{\alpha} |C_{\alpha}^{\text{ini}} |^4\!=\!\text{IPR}_{\text{ini}}^{-1} < 3/{\cal D}$. After reaching this point, $F(t)$ fluctuates around $\overline{F}$, the fluctuations decreasing exponentially with $L$ \cite{Zangara2013}. The inverse participation ratio~\cite{Izrailev1990,ZelevinskyRep1996}, $\text{IPR}_{\text{ini}} $, measures the level of delocalization of $|\text{ini}\rangle $ in the energy eigenbasis. Better filling of the energy shell necessarily implies larger $\text{IPR}_{\text{ini}} $. For $ |\rm{NS}\rangle$, as expected from Fig.~\ref{fig:shell}, $\overline{F}$ decreases significantly from integrable to chaotic Hamiltonians (see Fig.~\ref{fig:fidelity}), whereas for $|\rm{DW}\rangle$ and $|\rm{PS}\rangle$, $\overline{F}$ is actually largest for the chaotic $\hat{H}_{\Delta=1,\lambda=1}$, where $E_{\text{ini}}$ is closest to the border of the spectrum. Thus, the width of the energy shell and the extent of its filling are more important  in the characterization of the relaxation process than the actual regime of $\hat{H}_F$.
\begin{figure}[h!]
\centering
\includegraphics[width=0.45\textwidth]{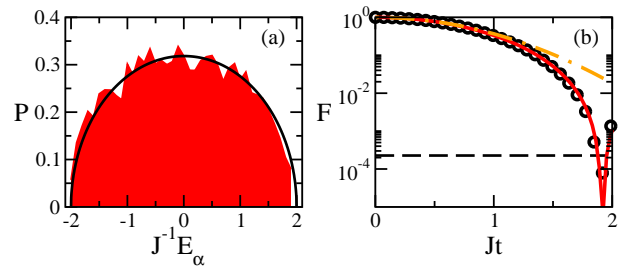}
\caption{(Color online) Energy distribution of a generic initial state (a) and fidelity decay (b) for full random matrices from a Gaussian Orthogonal Ensemble~\cite{Brody1981}; ${\cal D} \!=\! 12870$. The distribution is normalized so that $\sigma_{\rm{ini}} \!=\!{\cal E}/2\!=\!J$. Solid lines: $P^{\text{ini}}_{\text{FRM}}(E)$ (a) and $F_{\text{FRM}}(t)$ (b); dot-dashed: Eq.~(\ref{eq:fidelity}); dashed: saturation; circles: numerical data.}
\label{fig:RMT}
\end{figure}

We verified that the Gaussian decay of $F(t)$ until saturation is valid for initial states from different $\hat{H}_I$, such as XX and XXZ models~\cite{preparation}, and for different $\hat{H}_F$, such as disordered models~\cite{preparationTorres}, provided $P^{\text{ini}}_{\alpha}$ be Gaussian. Here  $t_R=\sqrt{\ln(\text{IPR}_{\text{ini}}) }/\sigma_{\text{ini}}$ sets the minimum time for an initial state with a single-peaked energy distribution and evolving under a two-body-interaction Hamiltonian to relax to equilibrium.

The fidelity decay can however be faster when more-body interactions are included. To identify the ultimate bound for the fidelity decay in the general scenario of strong perturbations, we resort to full random matrices. They do not describe realistic systems, because they imply the simultaneous interactions of many particles, but they provide the extreme case. The maximum LDOS in two-body-interaction Hamiltonians is Gaussian, reflecting these systems' density of states. The energy distribution of initial states evolving under full random matrices is broader, showing a semicircular shape, $P^{\text{ini}}_{\text{FRM}}(E)=\frac{2}{\pi {\cal E}} \sqrt{1 - \left(\frac{E}{{\cal E}}\right)^2}$, where  $2{\cal E}$ is the length of the spectrum [Fig.\ref{fig:RMT} (a)]. This reflects the semicircular form of the density of states of full random matrices~\cite{Brody1981}. The Fourier transform of the semicircular LDOS leads to the following fidelity decay: 
\begin{equation}
F_{\text{FRM}}(t) =  [{\cal J}_1( 2 \sigma_{\text{ini}} t)]^2/(\sigma_{\text{ini}}^2 t^2),
\end{equation}
 where  ${\cal J}_1$ is the Bessel function of the first kind. In this case, the fidelity behavior is clearly faster than Gaussian, as made explicit by the comparison between the two shown in Fig.~\ref{fig:RMT} (b) \cite{noteCos}.

%%%%%%%%%%%%%%%% OBSERVABLES %%%%%%%%%%%%%%%%%%%%%%%%%

\section{FEW-BODY OBSERVABLES}
Many factors come into play when describing the dynamics of few-body observables $\hat{A}$. Nevertheless, a simple general picture becomes viable  when $[\hat{A}, \hat{H}_I]=0$ (we recall that  $\hat{H}_I$ defines the basis and $|\text{ini}\rangle$ is one of the basis vectors). In this case, the evolution depends on the fidelity:
$A(t) = F(t) A(0) + \sum_{n\neq |\text{ini} \rangle} | \langle n| e^{-i \hat{H}_F t}|\text{ini}\rangle|^2 A_{nn} $, where $A_{nn} = \langle n |\hat{A} |n \rangle$.
For the $\hat{H}_I$ considered here, this implies observables in the $z$ direction, whereas for systems quenched from the XX model, this would mean $\hat{A}$ in the $xy$ plane, such as the kinetic energy.  
One can then infer from the expansion,
\begin{equation}
A(t) \!
 \simeq \! \left( 1 \!-\! \sigma_{\text{ini}}^2 t^2 \right) \! A(0) + \sum_{n\neq \text{ini}}   |\langle n |\hat{H}_F |\text{ini} \rangle|^2 t^2  A_{nn} ,
\label{eq:Oexpansion}
\end{equation}
that  any observable will show a very similar short-time dynamics for $\hat{H}_F$ inducing comparable squared off-diagonal elements $|\langle n |\hat{H}_F |\text{ini} \rangle|^2$, and thus similar $\sigma_{\text{ini}}$, even when the Hamiltonians have very different properties.

When $|\text{ini}\rangle$ is a site-basis vector and $\hat{A}$ is in a direction perpendicular to $z$, $A(0)\!=\!0$, so $F(t)$ does not play a part in the evolution. In this case, if $A_{nn'}$ is imaginary, the dominant terms in $A(t)$ are ${\cal O}(t)$ and similar dynamics emerges for distinct $\hat{H}_F$ with comparable $\langle n |\hat{H}_F |\text{ini} \rangle$ \cite{preparation}. This happens {\em e.g.} for the spin-current~\cite{Santos2011,Zotos1997}. If $A_{nn'}$ is real,  the dominant terms are ${\cal O}(t^2)$ and contributions from the diagonal elements of $\hat{H}_F$ (thus $\Delta$) complicate the picture. However, even for observables where $[\hat{A}, \hat{H}_I]\neq 0$, one can always construct particular $|\text{ini}\rangle$'s 
that lead to expressions equivalent to Eq.~(\ref{eq:Oexpansion}). As an example, we consider the spin-spin correlation in $x$ between sites $\frac{L}{2},\frac{L}{2}+1$ for a Bell-type state $|\text{BS}\rangle\!=\! (|\ldots\uparrow_{L/2} \downarrow_{L/2+1} \ldots \rangle \!+\! |\ldots\downarrow_{L/2} \uparrow_{L/2+1} \ldots \rangle )/\sqrt{2}$, where apart from  $\frac{L}{2},\frac{L}{2}+1$ the two bases are equal.

\begin{figure}[htb]
\centering
%\vskip 0.25 cm
\includegraphics[width=0.38\textwidth]{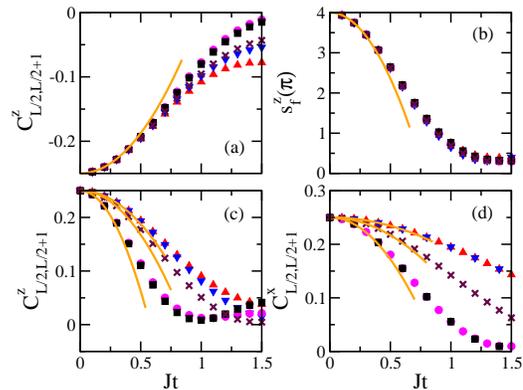}
\caption{(Color online) $C^{z,x}_{L/2, L/2+1}$ and $s^z_f(\pi)$ for $ |\rm{NS}\rangle$ (top), $ |\rm{PS}\rangle$ (c), and $ |\rm{BS}\rangle$ (d); $\hat{H}_F$ as in Fig.~\ref{fig:fidelity}; $L=16$. Solid lines give Eq.~(\ref{eq:Oexpansion}).}
\label{fig:ONeel}
\end{figure}
Figure~\ref{fig:ONeel} shows the spin-spin correlation in $z(x)$, $\hat{C}^{z(x)}_{L/2,L/2+1} \!\!=\!\! \hat{S}_{L/2 }^{z(x)} \hat{S}_{L/2+1}^{z(x)}$, and the structure factor in  $z$,
$\hat{s}_f^{z}(k) \!=\! \sum_{l,j=1}^L e^{i k (l-j) } \hat{S}_l^{z} \hat{S}_{j}^{z}/L$, with momentum $k$.  The first observable is local and the second is nonlocal in position. Despite their differences, for the initial states considered in the figure, their short-time dynamics is dictated by Eq.~(\ref{eq:Oexpansion}).

The top panels of Fig.~\ref{fig:ONeel} depict results for $|\rm{NS}\rangle$. For short-times, $\hat{C}^{z}$ and $\hat{s}_f^z$ are insensitive to the differences between the five $\hat{H}_F$, the curves almost coinciding. Strikingly, this behavior persists for $s_f^{z}(\pi)$ until the steady state. For $\hat{C}^{z}$, the saturation points  follow the filling of the energy shell: correlations remain for the integrable Hamiltonians, being larger for $\hat{H}_{\Delta=1,\lambda=0}$, and get close to zero for strongly chaotic cases. 

The bottom panels of Fig.~\ref{fig:ONeel} give results for $\hat{C}^{z}$ [$ |\rm{PS}\rangle$ in (c)] and $\hat{C}^{x}$ [$ |\rm{BS}\rangle$ in (d)]. For both observables the dynamics is very similar for Hamiltonians leading to the same $ \sigma_{\text{ini}}$.  Other observables and initial states reiterate these findings~\cite{preparation}.

%%%%%%%%%%%%%%%%%%%% CONCLUSION %%%%%%%%%%%%%%%%%%%%%

\section{CONCLUSIONS}
We studied the fidelity decay and the evolution of few-body observables in isolated interacting quantum systems initially far from equilibrium. The Hamiltonians dictating the evolution, the initial states $|{\text{ini}} \rangle $, and the observables analyzed are accessible to experiments with optical lattices and with nuclear magnetic resonance~\cite{Franzoni,Ramanathan2011}. 

We revised the commonly accepted picture that in these systems the fidelity decay had to transition to an exponential behavior before saturation. We demonstrated that when the energy distribution of $|{\text{ini}} \rangle $ is Gaussian, substantially filling the energy shell, the numerical results for the {\em fidelity decay can be Gaussian until saturation} and show excellent agreement with the analytical expression. This behavior sets the lower bound for the fidelity decay of isolated systems with two-body interactions and $|{\text{ini}} \rangle $ consisting of a single peaked energy distribution. We also derived an expression for the fidelity decay of full random matrices, which establishes the ultimate bound for quenched systems with many-body interactions.

The width $\sigma_{\text{ini}}$  plays a major role in the short-time dynamics of few-body observables that commute with $\hat{H}_I$. We thus find {\em observables showing a remarkable similar behavior} despite evolving according to  very different final Hamiltonians (distinct anisotropies and regimes).

The general features unveiled here for fidelity and observables constitute crucial steps towards a complete description of the relaxation process in realistic quantum systems of interacting particles. Our results are also key for quantum control methods~\cite{Bhattacharyya1983,Pfeifer1993,GiovannettiALL,Caneva2009} and quantum thermodynamics due to the link between LDOS and work distribution function~\cite{work}.

%%%%%%%%%%%%%%%% ACKNOWLEDGMENTS %%%%%%%%%%%%%%%%%%%%%
\section{ACKNOWLEDGMENTS}
This work was supported by the  NSF grant No.~DMR-1147430. E.J.T.H. acknowledges partial support from CONACyT, Mexico. We thank Felix Izrailev, Eduardo Mascarenhas, and Simone Montangero for discussions.

%%%%%%%%%%%%%%%%%%%% REFERENCES %%%%%%%%%%%%%%%%%%%%%

\end{document}